\documentclass[review]{elsarticle}

\usepackage{lineno,hyperref}
\modulolinenumbers[5]
\usepackage{amsmath}
\usepackage{amsfonts}
  \usepackage{paralist}
  \usepackage{color}
  \usepackage{graphics} 
  \usepackage{epsfig} 
\usepackage{graphicx}  
\usepackage{epstopdf}
\usepackage{tikz-cd}
\usepackage{mathrsfs}

\usepackage{tikz}
\usetikzlibrary{decorations.pathmorphing,patterns}

\newcommand{\F}{\mathbb{F}}
\newcommand{\R}{\mathbb{R}}

\newcommand{\proa}{A^*G \mbox{$\;$}_{\tau^*} \kern-3pt\times_\alpha
G \mbox{$\;$}_\beta \kern-3pt\times_{\tau^*} A^*G}

\newtheorem{theorem}{Theorem}[section]

\newtheorem{lemma}[theorem]{Lemma}

\newtheorem{proposition}{Proposition}

\newtheorem{definition}[theorem]{Definition}
\newtheorem{remark}{Remark}
\newtheorem{example}{Example}
\journal{Nonlinear Analysis: Hybrid Systems}
\begin{document}

\begin{frontmatter}

\title{
Symmetries and periodic orbits in simple hybrid Routhian systems\tnoteref{mytitlenote}}
\tnotetext[mytitlenote]{This work was partially supported by I-Link Project (Ref: linkA20079) from CSIC and by Ministerio de Econom\'ia, Industria y Competitividad (MINEICO, Spain) under grant MTM2016-76702-P. The project that gave rise to these
results received the support of a fellowship from ``la Caixa' Foundation'' (ID 100010434). The fellowship
code is LCF/BQ/PI19/11690016. The graduate student M.E. Eyrea Iraz\'u was partially supported by CONICET Argentina. L. Colombo and M.E. Eyrea Iraz\'u were partially supported by ``Severo
Ochoa Programme for Centres of Excellence'' in R$\&$D (SEV-2015-0554). The authors are indebted with the reviewers and editor for their recommendations that helped to improve the quality, clarity and exposition of this work.}

\author{Leonardo J. Colombo\fnref{1}}
\address{Instituto de Ciencias Matem\'aticas, Consejo Superior de Investigaciones Cient\'ificas}
\address{Calle Nicol\'as Cabrera 13-15, Cantoblanco, 28049, Madrid, Spain}
\fntext[myfootnote]{Email address: leo.colombo@icmat.es.}

\author{Mar\'ia Emma Eyrea Iraz\'u\fnref{1}\fnref{*}}
\address{Department of Mathematics, Universidad Nacional de La Plata}
\address{Calle 1 y 115, La Plata 1900, Buenos Aires, Argentina}
\fntext[*]{First and Corresponding author. Email address: maemma@mate.unlp.edu.ar.}

\begin{abstract}
Symmetries are ubiquitous in a wide range of nonlinear systems. Particularly in systems whose dynamics is determined by a Lagrangian or Hamiltonian function. For hybrid systems which possess a continuous-time dynamics determined by a Lagrangian function, with a cyclic variable, the degrees of freedom for the corresponding hybrid Lagrangian system can be reduced by means of a method known as \textit{hybrid Routhian reduction}. In this paper we study sufficient conditions for the existence of periodic orbits in hybrid Routhian systems which also exhibit a time-reversal symmetry. Likewise, we explore some stability aspects of such orbits through the characterization of the eigenvalues for the corresponding linearized Poincar\'e map. Finally, we apply the results to find periodic solutions in underactuated hybrid Routhian control systems. 
\end{abstract}


\end{frontmatter}

\section{Introduction}
Hybrid systems are dynamical systems with continuous-time and discrete-time components in its dynamics. These dynamical systems  are capable of modeling several physical systems, such as, multiple UAV systems \cite{lee}, \cite{sh2}, bipedal robots \cite{si}, \cite{sp}, \cite{sh}, embedded computer systems \cite{cog}, \cite{dimos}, \cite{hybridbook} and underactuated mobile vehicles \cite{frazzoli}, among others. 

Simple hybrid systems are a type of hybrid systems introduced in \cite{SHS}, called in this manner because of its simple nature. A simple hybrid system is characterized by a tuple $\mathbf{H}=(D, X, \mathcal{S}, \Delta)$ where $D$ is a smooth manifold, $X$ is a smooth vector field on $D$, $\mathcal{S}$ is an embedded submanifold of $D$ with co-dimension $1$ called the switching surface (or the guard), and $\Delta:\mathcal{S}\to D$ is a smooth embedding called the impact map (or the reset map). This type of hybrid system has been mainly employed for the understanding of walking gaits in bipeds and insects \cite{AmGrSp}, \cite{HoFuKoGu}, \cite{Biped-book}. In the situation where the vector field $X$ is associated with a mechanical system (Lagrangian or Hamiltonian), alternative approaches for mechanical systems with unilateral constraints have been considered in \cite{cortes}, \cite{cortes2}, \cite{ibort}, \cite{ibort2}, and \cite{LC}.

A symmetry is a transformation that leave invariant the solutions in a dynamical system. The type of symmetry that most mechanical systems naturally exhibit is known as reversing symmetry. Reversing symmetries leave the equations of motion invariant if the direction of time is reversed.  Dynamical systems possessing this class of symmetry are called reversible if the reversing symmetry is an involution \cite{Lamb}.  It is important to mention that the existence of periodic orbits has been a predominant topic of research in dynamical systems since the studies of Poincar\'e \cite{Po}. The use of reversing symmetries to find periodic orbits has been employed, for instance, in the restricted three-body problem \cite{bir}.

When a dynamical system exhibits a symmetry, it produces a conserved quantity for the system. This reduces the degrees of freedom in the dynamics of the system. One of the classical reduction by symmetry procedures in mechanics is the Routh reduction method \cite{Goldstein}. During the last few years there has been a growing interest in Routh reduction, mainly motivated by physical applications \cite{santiago}, \cite{GT2}, \cite{Babo}, \cite{GT}.  Routh reduction for hybrid systems has been introduced by A. Ames and S. Satry and it has been applied in the field of bipedal locomotion \cite{amesthesis}, \cite{AmGrSp}, \cite{amesrouth}. The reduced simple hybrid system is called simple hybrid Routhian system. In this work we build in the former approach to that concept by studying sufficient conditions for which a simple hybrid Routhian system exhibits a periodic solution.

The search of limit cycles in hybrid systems has been an active research field in the robotics and automatic control community since the works of Mc'Geer due to the study of periodic walking gaits for passive dynamic walkers \cite{ruina}, \cite{mc}. Since these works, the study of orbital stability for hybrid systems has been the more explored analysis in this field.  The method of Poincar\'e map is frequently used in the legged locomotion community to study orbital stability of walking gaits \cite{grizzle}, \cite{KH}, \cite{SVF}, \cite{hamed}, \cite{wendel}, \cite{Biped-book}. In most of the studies analyzed in the literature employing such an approach, one assumes the existence of a periodic solution. Then one proceed with the corresponding stability analysis of the orbits by examining the eigenvalues of the linearized Poincar\'e Map at its fixed points. In general, since fixed points of the Poincar\'e map corresponds with periodic orbits for the underlying dynamical system, to find these orbits one needs to employ computational resources to find the Poincar\'e map and its fixed points.

In this work, we show how to ensure the existence of periodic solutions by examining the symmetries for a simple hybrid Routhian system excluding computational tools.  Moreover, with the method proposed in this work we also provide a characterization for the eigenvalues associated with the linearized Poincar\'e map for these periodic solutions, also from an analytical point of view. To the best of our knowledge, sufficient conditions for simple hybrid Routhian systems under which one can ensure the existence of periodic solutions and the characterization of its qualitative behavior, excluding the use of computational tools, has not yet been widely discussed in the literature. Similar results for simple hybrid (non-Routhian) systems concerning the existence of periodic solutions can be found in \cite{BlClCo} and \cite{CoClBl}. Sufficient conditions for the existence and uniqueness of Poincar\'e maps can be found in \cite{ClBlCo} and results about stability of periodic solutions for 2D-simple hybrid systems can be found in \cite{GoCo}.

The primary goal of this paper consists on establishing the conditions under which we can  ensure the existence of periodic orbits in simple hybrid Routhian systems. By introducing a time-reversal symmetry, trajectories for a simple hybrid Routhian system will become in a periodic orbit if the trajectory begins at a fixed point of the symmetry map. In this work, we provide characterizations for the eigenvalues associated with the linearized Poincar\'e map for these periodic solutions. We also apply the results to the classical example of the 2D one-leg robotic hopper for which, after applying Routh reduction method for simple hybrid system, we reduce the hybrid dynamics to the one for the 2D spring loaded inverted pendulum (SLIP). Then we study the existence of periodic motions and show that periodic orbits are at most marginally stable. As an application we employ the results given in this work for the search of periodic orbits in underactuated hybrid mechanical control systems, in particular, for the 2D spring loaded inverted pendulum.
  
The paper is structured as follows: Section $2$ presents a review on Lagrangian mechanics and Routh reduction. Section $3$ introduces simple hybrid Routhian systems and the 2D-SLIP model is derived as the simple hybrid Routhian system for to the 2D one leg robotic hopper. Section $4$ contains the main results of the paper, such as how after introducing time-reversal symmetries we find sufficient conditions to ensure the existence of periodic solutions and we study how to characterize some of the eigenvalues for the linearization of the Poincar\'e map corresponding to the periodic orbit. In this section we also explain how to apply the results for the reduced system associated with the 2D one leg robotic hopper. Finally, Section $5$ shows the results obtained in Section $4$ to study the existence of periodic solutions in underactuated simple hybrid Routhian control systems.

\section{Preliminaries on Routh reduction}\label{Sec:2}
Let $Q$ be the configuration space of a mechanical system, a differentiable manifold of dimension $n$, with local coordinates $q=(q^1,\ldots,q^n)$. Let $TQ$ be the tangent bundle of $Q$, locally described by the positions and velocities for the system $(q,\dot{q})=(q^1,\ldots, q^n,\dot{q}^{1},\ldots,\dot{q}^{n})\in TQ$ with $\hbox{dim}(TQ)=2n$. 

The dynamics of the mechanical system is determined by a Lagrangian function $L:TQ\to\mathbb{R}$ given by $L(q,\dot{q})=K(q,\dot{q})-V(q)$ where $K:TQ\to\mathbb{R}$ is the kinetic energy and $V:Q\to\mathbb{R}$ the potential energy. Along this work we will assume that the Lagrangian $L$ is regular, that is, $\displaystyle{\det\left(\frac{\partial^2L}{\partial\dot{q}^{i}\partial\dot{q}^{j}}\right)\neq0}$ for all $i,j=1,\ldots,n$.

The corresponding equations of motion describing the dynamics of the system are given by the Euler-Lagrange equations for $L$, that is, $\displaystyle{\frac{d}{dt}\left(\frac{\partial L}{\partial\dot{q}^{i}}\right)=\frac{\partial L}{\partial q^{i}}}$, $i=1,\ldots,n;$ a system of $n$ second-order ordinary differential equations. These equations induce a vector field $X_L:TQ\to T(TQ)$ describing the dynamics of the Lagrangian system, given by $$X_L(q^i,\dot{q}^i)=\left(q^i,\dot{q}^i;\dot{q}^i, \left(\frac{\partial^2L}{\partial\dot{q}^{i}\partial\dot{q}^{j}}\right)^{-1}\left(\frac{\partial L}{\partial q^i}-\frac{\partial^2L}{\partial\dot{q}^{i}\partial q^j}\dot{q}^{j}\right)\right).$$

There exists a large class of systems for which the Lagrangian does not depend on some of the generalized coordinates. Such coordinates are called \emph{cyclic}  or \emph{ignorable}, and the corresponding generalized momenta is easily checked to be constants of the motion.

 The Routh reduction procedure is a classical reduction technique which takes advantage of the conservation law to define a \textit{reduced Lagrangian} function, so-called the \emph{Routhian}, such that the solutions of the Euler-Lagrange equations for the Routhian are in correspondence with the solutions of Euler-Lagrange equations for the original Lagrangian, when the conservation of momenta is taken into account. The technique is due to Edward Routh, who successfully applied it to the study of the stability of steady motions (today, we call these relative equilibria).

The steps to carry out Routh reduction are described in many classical texts of mechanics (such as \cite{Goldstein}) as follows.  Assume $q^{i}=(x^{a},\theta)$ are local coordinates on $Q$, $a=1,\ldots,n-1$.
\begin{enumerate}[1)]
\item Let $L(x^a,\dot x^a,\theta,\dot\theta)$ be a $G$-regular Lagrangian with cyclic coordinate $\theta$, that is, $\displaystyle{\frac{\partial L}{\partial\theta}=0}$, and denote by $p_\theta$ the generalized momentum corresponding to $\theta$.

\item Fix a value of the momentum $\mu=p_\theta$, and consider the function
\begin{equation}\label{eq:Routh}
R_c^\mu(x^a,\dot x^a)=\left(L-\dot\theta p_\theta\right)\Big{|}_{p_\theta=\mu}\,,
\end{equation}where the notation means that we have used the relation $\mu=p_\theta$ to replace all the appearances of $\dot\theta$ in terms of $(x^a,\dot x^a)$ and the parameter $\mu$. The function $R_c^\mu$ is the (\emph{classical}) \textit{Routhian}.

\item If we regard $R_c^\mu$ as a new Lagrangian in the variables $(x^a,\dot x^a)$, then the solutions of the Euler-Lagrange equations for $R_c^\mu$ are in correspondence with those of $L$ when one takes into account the relation $p_\theta=\mu$. More precisely, if $L$ is $G$-regular:

    \begin{itemize}
    \item[(i)] Any solution $(x^a(t),\theta(t))$ of the Euler-Lagrange equations for $L$ with momentum $p_\theta=\mu$ projects onto a solutions $x^a(t)$ of the Euler-Lagrange equations for $R_c^\mu$, \begin{equation}\label{routh1eq}\frac{d}{dt}\left(\frac{\partial R_c^{\mu}}{\partial\dot{x}^{a}}\right)-\frac{\partial R_c^{\mu}}{\partial x^{a}}=0.\end{equation} These equations will be referred to as \textit{Routh equations}. Equations \eqref{routh1eq} induce a vector field $X_{R_c^{\mu}}:TQ\to T(TQ)$ describing the dynamics of the reduced system, called \textit{Routhian vector field}, and given by $$X_{R_c^{\mu}}(x^a,\dot{x}^a)=\left(x^a,\dot{x}^a,\dot{x}^a, \mathcal{M}_{ab}\left(\frac{\partial L}{\partial x^a}-\frac{\partial^2L}{\partial\dot{x}^{a}\partial x^b}\dot{x}^{b}\right)\right),$$ where $\mathcal{M}_{ab}=\left(\frac{\partial^2L}{\partial\dot{x}^{a}\partial\dot{x}^{b}}\right)^{-1}$.
    \item[(ii)] Conversely, any solution of Routh equations for $R^\mu_c$ can be lifted to a solution of the Euler-Lagrange equations for $L$ with momentum $p_\theta=\mu$.
    \end{itemize}
\end{enumerate}

This is best understood by means of an example:

\begin{example} The Lagrangian $L:T(\mathbb{R}\times\mathbb{S}^{1})\to\mathbb{R}$ given by \[
L(r,\dot r,\theta,\dot\theta)=\frac{1}{2}m(\dot r^2+r^2\dot\theta^2)-\frac{1}{2}r^2k.
\] describes (in polar coordinates) the motion of a mass $m$ on the plane which is pinned to a fixed point through a spring of elastic constant $k$ (as it is shown in Figure \ref{fig1}).
\begin{figure}[h!]\label{fig1}\centering
 \includegraphics[width=4.5cm]{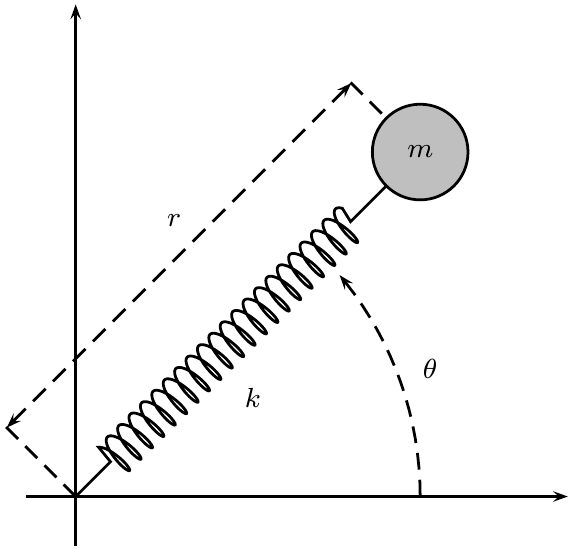}
     \caption{spring-loaded pendulum}
\end{figure}

Note that the Lagrangian $L$ is regular since $\displaystyle{\det\begin{pmatrix} \frac{\partial^2 L}{\partial\dot{r}\partial\dot{r}} & \frac{\partial^2 L}{\partial\dot{r}\partial\dot{\theta}} \\ \frac{\partial^2 L}{\partial\dot{\theta}\partial\dot{r}} & \frac{\partial^2 L}{\partial\dot{\theta}\partial\dot{\theta}} \end{pmatrix}=(mr)^2\neq0.}$ The coordinate $\theta$ is cyclic, and the associated conservation law reads $p_\theta=mr^2\dot \theta$. If we set $p_\theta=\mu\neq 0$ (a fixed constant $\mu$ which represents the fixed value of the momentum), we can work out $\dot\theta$ in terms of the $(r,\dot r)$ and find the relation $\dot \theta = \mu/mr^2$. Then the Routhian $R_c^{\mu}:T\mathbb{R}\to\mathbb{R}$ is given by 
\[
R_{c}^\mu(r,\dot r) = \left(L-\mu\dot\theta\right)\Big{|}_{p_\theta=\mu}= \frac{1}{2}\left(m\dot r^2-kr^2-\frac{\mu^2}{mr^2} \right). 
\]
The equivalence of solutions between $L$ and $R_{c}^\mu$ happens when one takes into account the conservation law $p_\theta=\mu$. More precisely: a solution $(r(t),\theta(t))$ of the Euler Lagrange for  the $G$-regular Lagrangian $L$ with momentum $p_\theta=\mu$ projects onto a solution $r(t)$ of Routh equations for $R_{c}^\mu$. Conversely, given a solution $r(t)$ of the Routh equations for $R_{c}^\mu$, one can use the conservation law $p_\theta=\mu$ to lift $r(t)$ to a solution 
$(r(t),\theta(t))$ of the Euler Lagrange for $L$ (with momentum $\mu$, obviously). \end{example}

\section{Simple hybrid Routhian systems}\label{Sec3.A}
\subsection{Simple hybrid system}
Simple hybrid systems \cite{SHS} (see also \cite{amesthesis}) are characterized by the 4-tuple $\mathbf{H}=(D, X, \mathcal{S}, \Delta)$ where $D$ is a smooth manifold, the \textit{domain}, $X$ is a smooth \textit{vector field} on $D$, $\mathcal{S}$ is an embedded submanifold of $D$ with co-dimension $1$ called \textit{switching surface}, and $\Delta:\mathcal{S}\to D$ is a smooth embedding called the \textit{impact map}. $\mathcal{S}$ and $\Delta$ are also called the \textit{guard} and \textit{reset map}, respectively, in \cite{amesthesis}-\cite{amesham}.

In this work, the dynamics associated with a simple hybrid system is described by an autonomous system with impulse effects as in \cite{Biped-book}. We denote by $\Sigma_{\textbf{H}}$ the \textit{simple hybrid dynamical system} generated by $\mathbf{H}$, that is,  \begin{equation}\label{LHS}\Sigma_{\textbf{H}}:\begin{cases} \dot{\gamma}(t)=X(\gamma(t)),\quad\quad \gamma^{-}(t)\notin\mathcal{S} \\ \gamma^{+}(t)=\Delta(\gamma^{-}(t)),\quad \gamma^-(t)\in\mathcal{S} \end{cases}\end{equation} where $\gamma:I\subset\mathbb{R}\to D$, and $\gamma^{-}(t):=\displaystyle{\lim_{\tau\to t^{-}}}\gamma(\tau)$,  $\gamma^{+}(t):=\displaystyle{\lim_{\tau\to t^{+}}}\gamma(\tau)$ are the left and right limits of the state trajectory $\gamma(t)$, respectively, describing the states immediately before and after the times when integral curves of $X$ intersects $\mathcal{S}$ (i.e., pre and post impact of the solution $\gamma(t)$ with $\mathcal{S}$).

\begin{remark} Consider the impact map $\Delta$ given by the identity map. When a trajectory crosses $\mathcal{S}$, we will have $\gamma^+ = \Delta(\gamma^-) = \gamma^- \in \mathcal{S}$, so that we are again in the regime of discrete dynamics where re-initialization (to $\gamma^-$) will occur. It is clear that this process will never end. Consequently, there exists an infinite number of resets in a finite amount of time. This situation generates a class of behavior called \textit{Zeno behavior}.  That is, a solution of a hybrid system may experience a Zeno state if infinity many impacts occur in a finite amount of time \cite{amesthesis}. This is particularly problematic in applications where numerical work is conducted, as computation time grows infinitely large at these Zeno points.   There are two primary modes through which zeno behavior can occur:

$(1)$ A trajectory is reset back onto the guard, prompting additional resets. As seen in the above example where  $\gamma^+ = \Delta(\gamma^-) = \gamma^- \in \mathcal{S}$, if there is a set of points in the guard which the reset map cycles between, we can get 'stuck' on the guard. To exclude this type of behavior, we require that $S \cap \overline{\Delta}(S) = \emptyset$, where $\overline{\Delta}(S)$ denotes the closure as a set of $\Delta(S)$. This ensures that the trajectory will always be reset to a point with positive distance from the guard.

$(2)$ The set of times where a solution to our system reaches the guard (and is correspondingly reset) has a limit point. This happens, for example, in the case of the bouncing ball with coefficient of restitution $1/2$. If $t_0$ is the time between two impacts, then the time between the next two impacts will be $t_0/2$, then $t_0/4$, and so on. In time $$T = t_0 + \frac{t_0}{2} + \frac{t_0}{4} +... = 2t_0$$
we will have infinite resets in a finite time. To exclude these kind of situations, we require the set of impact times to be closed and discrete, as in \cite{Biped-book}.

 Both of the previously established hypothesis will be assumed implicitly throughout the rest of the paper (i.e.,  $\overline{\Delta}(\mathcal{S})\cap\mathcal{S}=\emptyset$ and the set of impact times is closed and discrete). Necessary and sufficient conditions for the existence of Zeno behavior in the class of hybrid systems studied in this work have been explored in  \cite{LA}.
\end{remark}

\subsection{Simple hybrid Routhian  systems}
When the configuration space is $Q=P\times G$ (i.e., $D=TQ=TP\times TG$) with $G$ an abelian Lie group and $P$ a smooth manifold, A. Ames and S. Sastry introduced the notion of hybrid Routhian  systems in \cite{amesthesis} and \cite{amesrouth}. This is based on 2 invariance properties allowing the reduction by symmetries: a cyclic regular Lagrangian $L:TQ\to\mathbb{R}$ and a function $h:Q\to\mathbb{R}$ describing an unilateral constraint which induces the switching surface, being $h$ cyclic and defined in the same generalized coordinates as $L$. These  two invariance properties allow to define the Routhian $R_c^{\mu}:TP\to\mathbb{R}$ and the reduced constraint function $\bar{h}:P\to\mathbb{R}$.

The starting point for symmetry reduction is a Lie group action $\psi\colon G\times Q\to Q$ of some Lie group $G$ on the manifold $Q$. There is a natural lift of the action $\psi$ to the space $T^*Q$, 
\begin{align*}
\Psi^{T^{*}Q}\colon  G\times T^*Q&\to T^*Q,\\
(g,(q,p))&\mapsto T^*\psi_{g^{-1}}(q,p).
\end{align*}
The cotangent bundle $T^{*}Q$ is equipped with the following geometric structure, called \textit{canonical symplectic structure} (see, for instance, \cite{Marsden1999}) $\Omega=dq\wedge dp$ with $(q,p)$ being local coordinates on $T^{*}Q$. The action $\Psi^{T^{*}Q}$ enjoys the following properties (see \cite{Marsden1999}):
\begin{itemize}
 \item $\Psi^{T^{*}Q}$ is a symplectic action, meaning that, if we denote $\Psi^{T^{*}Q}_g\equiv \Psi^{T^{*}Q}(g,\cdot)$, $(\Psi^{T^{*}Q}_g)^*\Omega=\Omega$, where $(\Psi^{T^{*}Q}_g)^*\Omega$ accounts for the pullback by $\Psi^{T^{*}Q}_g$ of the $2$-form $\Omega$.
 \item It admits an $\hbox{Ad}^*$-\textit{equivariant momentum map}  $J\colon  T^*Q\to \mathfrak{g}^{*}$ given by
 \[
  \langle J(q,p), \xi\rangle=\langle p, \xi_Q\rangle,\quad \forall \xi\in \mathfrak{g}^{*},
 \]
 where $\xi_Q(q):=\frac{d}{dt}\psi_{\exp(t\xi)}q$ is the \textit{infinitesimal generator} of the element $\xi\in \mathfrak{g}^{*}$ and where $\mathfrak{g}^{*}$ denotes the dual of $\mathfrak{g}$, the Lie algebra associated with the Lie group $G$.
\end{itemize}
Likewise, there is a lift of the action $\psi$ to $TQ$ denoted by $\Psi^{TQ}$
\begin{align*}
\Psi^{TQ}\colon  G\times TQ&\to TQ,\\
(g,(q,\dot q))&\mapsto T\psi_{g}(q,\dot q).
\end{align*}

It has been shown in \cite{amesthesis} and \cite{amesrouth} that to perform a hybrid Routhian reduction one needs to impose some compatibility conditions between the action and the simple hybrid system whose continuous-time dynamics is described by a Lagrangian flow associated to $L$ (a.k.a simple hybrid Lagrangian system). By the term \emph{hybrid action} we mean a Lie group action $\psi\colon G\times Q\to Q$ such that (i) $L$ is invariant under $\Psi^{TQ}$, i.e. $L\circ \Psi^{TQ}=L$; (ii) $\Psi^{TQ}$ restricts to an action on $\mathcal{S}$; (iii) $\Delta$ is equivariant w.r.t. the previous action, namely  $\Delta\circ \Psi^{TQ}_g\mid_{\mathcal{S}}=\Psi^{TQ}_g\circ \Delta$.

In the case of a hybrid action, $\Psi^{TQ}$ admits an $\hbox{Ad}^*$-equivariant (Lagrangian) momentum map $J_L: TQ\to\mathfrak{g}^{*}$ given by $J_L=J\circ\F L$, where $\mathbb{F}L:T\to T^{*}Q$ is the Legendre transformation associated with regular Lagrangian $L$, given by $\mathbb{F}L(q,\dot{q}):=(q,p=\partial L/\partial\dot{q})$. This follows directly from the invariance of $L$, since it implies that $\F L$ is an equivariant diffeomorphism, i.e. $
\F L\circ \Psi^{TQ}_g=\Psi_g^{T^{*}Q}\circ \F L$.

The hybrid equivalent of momentum map is the notion of \emph{hybrid momentum map} introduced in   \cite{amesthesis}. For example, in the case of $TQ$, $J_L$ is a  \emph{hybrid momentum map} if the  diagram
\begin{equation}\label{diag1}
\begin{tikzcd}[column sep=1.5cm, row
    sep=1.2cm]
& \mathfrak{g}^{*} &\\
TQ \arrow[ur,"J_L"] & \mathcal{S} \arrow[u,"J_L\mid_{\mathcal{S}}"] \arrow[l,swap,hook',"i"] \arrow[r,"\Delta"] & TQ  \arrow[ul, swap,"J_L"]
\end{tikzcd} 
\end{equation}
commutes, where $i$ denotes the canonical inclusion from $\mathcal{S}$ to $TQ$.

The situation of interest in this paper is that of an Abelian group action, i.e. $G=\mathbb{S}^1$ (the case $G=\R$ is analogous; and  if $G$ is a product one can iterate the procedure): this corresponds to the classical notion of cyclic coordinates. From now on we will assume $Q=P\times\mathbb{S}^1$ where $P$ is called the \emph{shape space} and the action being given by
\begin{align}\label{eq:action}
\nonumber \psi_{\alpha}\colon \mathbb{S}^1\times Q&\to \mathbb{S}^1\times Q ,\\
(\theta,x)&\mapsto (\theta+\alpha,x).
\end{align}
While this is indeed a strong assumption, it is always the case locally, so as long as it applies to the domain of interest of a specific problem the procedure below applies. More general results where the manifold is not a product or the Lie group is arbitrary (not necessarily Abelian) can be handled by using the same tools, but involving more technicalities such as the introduction of a principal connection. These matters will be discussed in a future extension of this work.

A. Ames and S. Sastry also shown that if the trajectory $\gamma$ for a simple hybrid Lagrangian system starting at $\gamma_0\in J^{-1}(\mu)$, with $J:TQ\to\mathfrak{g}^{*}$ the momentum map associated for the conserved quantity $\mu$, being $\mathfrak{g}$ the Lie algebra associated with $G$, then the trajectory for a simple hybrid Routhian system starting at $\pi(\gamma_0)$ with $\pi:TQ\to TP$ the projection over the first factor of $TQ=TP\times TG$, is determined by $\pi(\gamma(t))$.

\begin{definition}\label{def1}
A simple hybrid system $\mathbf{H}=(D, X, \mathcal{S}, \Delta)$ is said to be a \textit{simple hybrid Routhian system} if it is determined by $\mathbf{H}^{R_c^{\mu}}:=(TP, X_{R_c^{\mu}}, \mathcal{S}^{\mu}, \Delta^{\mu})$, where $X_{R_c^{\mu}}:TP\to T(TP)$ is the Routhian vector field, $\mathcal{S}^{\mu}$ is the  reduced switching surface given by $\mathcal{S}\mid_{(J_L\mid_{\mathcal{S}})^{-1}(\mu)}$ with $\mu$ the momentum constraint determined by the conserved quantity which arises from the cyclic coordinate and $\Delta^{\mu}:\mathcal{S}^{\mu}\to TP$ is the impact map on $\mathcal{S}^{\mu}$ given by $\Delta\mid_{(J_{L}\mid_{\mathcal{S}})^{-1}(\mu)}$.\end{definition}

\begin{definition}The \textit{simple hybrid Routhian dynamical system} generated by $\mathbf{H}^{R_c^{\mu}}$ is given by
 \begin{equation}\label{RHDS}\Sigma_{\mathbf{H}^{R^{\mu}_c}}:\begin{cases} \dot{\gamma}(t)=X_{R_c^{\mu}}(\gamma(t)), \hbox{ if } \gamma^{-}(t)\notin\mathcal{S}^{\mu},\\ \gamma^{+}(t)=\Delta^{\mu}(\gamma^{-}(t)),\hbox{ if } \gamma^-(t)\in\mathcal{S}^{\mu}, \end{cases}\end{equation}where $\gamma(t)\in TP $.\end{definition}
 
 Note that, as before, $\Delta^{\mu}:\mathcal{S}^{\mu}\to TP$ is continuous and if we denote $\overline{\Delta^{\mu}}(\mathcal{S}^{\mu})$ the closure of $\Delta^{\mu}(\mathcal{S}^{\mu})$ then we must assume $\overline{\Delta^{\mu}}(\mathcal{S}^{\mu})\cap \mathcal{S}^{\mu}=\emptyset$ and therefore, an impact does not lead immediately to another impact. We shall assume that $\textbf{H}^{R^{\mu}_c}$ satisfies  (see Section 4.1 in reference \cite{Biped-book} for more details)

\textbf{(A1) Assumption 1:} $\mathcal{S}^{\mu}\neq\emptyset$ and there exists an open subset $U\subset TP$ and a differentiable function $\bar{h}:U\to\mathbb{R}$ such that $\mathcal{S}^{\mu}=\{x\in U \mid \bar{h}(x)= 0\}$ with $\frac{\partial \bar{h}}{\partial x}(s)\neq 0$ for all $s\in\mathcal{S}^{\mu}$ (that is, $\mathcal{S}^{\mu}$ is an embedded submanifold of $TP$ with co-dimension $1$) and the Lie derivative of the vector field $X_{R_c^{\mu}}$ with respect to $\bar{h}$ does not vanish on $TP$, that is $L_{X_{R_c^{\mu}}}\bar{h}(w)\neq 0$, $\forall w\in TP$. 

\textbf{(A2) Assumption 2:} A trajectory $\gamma:[0,T]\to TP$ crosses the switching surface $\mathcal{S}^{\mu}$ at $t_{i}^{-}=\hbox{inf}\{t>0|\gamma(t)\in \mathcal{S}^{\mu}\}$. We allow the trajectory $\gamma$ to be non-smooth but continuous at $t_i^{-}$. That is, the velocity before the impact $\dot{x}^{-}$ may be different from the velocity $\dot{x}^{+}$ after the impact at $\mathcal{S}^{\mu}$, i.e., $\dot{x}(t_i^{-})\neq \dot{x}(t_i^{+})$.

The requirement that the configuration right after the impact does not belong to $\mathcal{S}^{\mu}$ (that is,  $\overline{\Delta^{\mu}}(\mathcal{S}^{\mu})\cap \mathcal{S}^{\mu}=\emptyset$), becomes a requirement on the exit velocity which states that the system has to be moving away from the switching surface right after the impact, that is, $\nabla \bar{h}(\gamma(t_i))\cdot \dot{\gamma}(t_{i}^{+})\leq 0,$ with $\bar{h}: TP\to\mathbb{R}$ as in \textbf{A1}.

A \textit{trajectory} of a hybrid Routhian system is determined by the dynamics associated with the Routhian until the instant when the state attains the switching surface $\mathcal{S}^{\mu}$. We refer to such instant as the \textit{impact time}. There is an instantaneous change in the velocity component of the state at impact times.  The impact map gives new initial conditions from which the continuous dynamics evolves until the next impact occurs.

\begin{remark}

We defined simple hybrid Routhian systems from simple hybrid systems as in \cite{amesthesis}-\cite{amesham}. However our definition is slightly different, but it is not contradictory, with the one given in \cite{amesthesis} and \cite{AmGrSp}. The constraint defining $\mathcal{S}^{\mu}$ is smooth satisfying both assumptions \textbf{A1} and \textbf{A2}, (it is not determined by unilateral constraints as considered in \cite{amesthesis}-\cite{amesham} and \cite{brogliato}, since we use the approach for systems with impulsive effects following \cite{grizzle} and  \cite{Biped-book}). Usually, and specially in biped locomotion, the impact map is given by foot placement on the ground and it comes from a Newtonian impact equation  \cite{AmGrSp}, \cite{amesham}, \cite{brogliato}. In this work we only assume $\Delta^{\mu}$ is a smooth embedding on $\mathcal{S}^{\mu}$ satisfying \textbf{A1} and \textbf{A2} .
\end{remark}

Solutions for the simple hybrid Routhian dynamical system $\Sigma_{\mathbf{H}^{R_c^{\mu}}}$, are considered right continuous and with finite left and right limits at each impact with $\mathcal{S}^{\mu}$. More precisely: 
\begin{definition}\label{defsol}
A \textit{solution} for the  hybrid Routhian system $\Sigma_{\textbf{H}^{R_c^{\mu}}}$ is a curve $\gamma:[t_0,t_f)\to TP$, $t_f\in\mathbb{R}\cup\{\infty\}$, $t_{f} > t_0$, unique from a given initial condition, depending continuously on it, satisfying \textbf{A2}, and such that: 
\begin{enumerate}
\item[(i)] $\gamma(t)$ is right continuous on $[t_0,t_f)$,
\item[(ii)] left and right limits,  denoted by $\gamma^{-}(t)$ and $\gamma^{+}(t)$, respectively, exists at each point $t\in(t_0,t_f)$,
\item[(iii)] there exists a closed discrete subset $\mathcal{I}\subset[t_0,t_f)$, the \textit{impact times}, such that, for each $t\notin\mathcal{I}$, $\gamma(t)$ is differentiable, $\dot{\gamma}(t)=X_{R_c^{\mu}}(\gamma(t))$, and $\gamma(t)\notin\mathcal{S}^{\mu}$; and for $t\in\mathcal{I}$, $\gamma^{-}(t)\in\mathcal{S}^{\mu}$ and  $\gamma^{+}(t)=\Delta^{\mu}(\gamma^{-}(t))$. 
\end{enumerate} 
\end{definition}

Note that right continuity of solutions implies $\gamma(t)=\gamma^{+}(t)$ at all the points in its domain of definition. If $\alpha_0\in TP$ denotes the initial state at time $t_0$, the solution at $t_0$ is denoted $\gamma(t_0,\alpha_0)$. When $\alpha_0\notin\mathcal{S}^{\mu}$, $\gamma(t_0,\alpha_0)=\alpha_0$ and when $\alpha_0\in\mathcal{S}^{\mu}$, $\gamma(t_0,\alpha_0)=\Delta^{\mu}(\alpha_0)=\gamma(t_0,\Delta^{\mu}(\alpha_0))$ (see Section $4.1$ in \cite{Biped-book} for details). 

\begin{remark}

As we commented in Section \ref{Sec:2}, since the Euler-Lagrange equations for $R^\mu_c$ involve less variables, they are easier to solve. If the Lagrangian $L$ is regular, then one usually proceeds to solve these first, and then uses the momentum constraint $\partial L/\partial \dot\theta=\mu$ to reconstruct the sought solutions of the Euler-Lagrange equations of $L$.
Special care should be taken when translating this reduction technique to hybrid systems. The reason is that the collisions with the switching surface will, in general, modify the value of the momentum map. Therefore, if $\mathcal{I}=\{I_{i}\}_{i\in \Lambda}$, where $\Lambda=\{0,1,2,...\}\subseteq \mathbb{N}$ is a finite (or infinite) indexing set, and $I_{i}=[t_{i},t_{i+1}]$ if $i, i+1\in \Lambda$ and $I_{N-1}=[t_{N-1},t_{N})$ or $[t_{N-1},\infty)$ if $|\Lambda|=N$, $N$ finite, with $t_{i},t_{i+1},t_{N-1}\in \R$ and $t_{i}\leq t_{i+1}$, then the Routhian has to be defined in each $I_i$ taken into account the value of the momentum $\mu_i$ after the collision at time $\tau_i$. Note that this also has influence in the way the reset map $\Delta$ is reduced. 

Let us denote: (1) $\mu_i$ the momentum of the system in $I_i=[t_i,t_{i+1}]$, (2) $\Delta^{\mu_i}$ is given by $\Delta\mid_{(J_{L}\mid_{\mathcal{S}})^{-1}(\mu_i)}$, and (3) $\mathcal{S}^{\mu_i}$ is given by $\mathcal{S}\mid_{(J_{L}\mid_{\mathcal{S}})^{-1}(\mu_i)}$, There is a sequence of hybrid Routhian systems

\[
\begin{tikzcd}[column sep=.5cm, row
    sep=.7cm]
 {[t_0,t_1]} \arrow[d,swap,"\text{Coll.}"]\arrow[r,"\text{Red.}"] & (P,R^{\mu_0}_c,\mathcal{S}^{\mu_0},\Delta^{\mu_0}) \arrow[d,"\text{Coll.}"]\\
 {[t_1,t_2]} \arrow[d,swap,"\text{Coll.}"]\arrow[r,"\text{Red.}"] & (P,R^{\mu_1}_c,\mathcal{S}^{\mu_1},\Delta^{\mu_1}) \arrow[d,"\text{Coll.}"]\\
(\dots) \arrow[r,"\text{Red.}"] & (\dots)
\end{tikzcd} 
\]

Similarly as in \cite{amesrouth}, the reconstruction procedure from the reduced hybrid flow to the flow for the simple hybrid Lagrangian system involves a recursive integration at each stage in the previous diagram of the cyclic variable using the solution of the reduced hybrid Routhian system. Roughly speaking, this accounts to imposing the momentum constraint on the reconstructed solution.

\end{remark}

\begin{example}[The 2D one leg robotic hopper]\label{slip}
The 2D one leg hopper robot consists of a spring loaded inverted pendulum together with a planar rigid body attached at the top of the spring (see Figure \ref{fig2}). This model is a schematic representation for the stance phase of a running or hopping biped with one foot on the ground at any time (see 
\cite{Al} for details).   The common point of attachment is the center of mass of the rigid body (i.e., the sprung leg is attached at a hip joint which is the center of mass).

\begin{figure}[h!]\label{fig2}\centering
 \includegraphics[width=4.5cm]{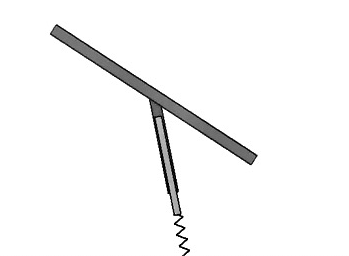}
     \caption{2D one leg robotic hopper}
\end{figure}

The configuration space of the system is $Q=(\mathbb{R}\times\mathbb{S}^{1})\times\mathbb{S}^{1}$, locally parametrized by the coordinates $q=(\xi,\varphi,\theta)$ describing the length of the spring, the angle of the spring with respect to the ground (i.e., the angle formed between the line joining the foothold to the center of mass and the vertical gravitational axis) and the attitude for the rigid body, respectively. We denote by $m$ the mass of the rigid body and $I$ its moment of inertia. The spring is considered massless and $l_0$ denotes the non-load length of the spring.

The motion is divided into two phases: The first one is the stance phase, with foothold fixed, the leg under compression, and  the body swinging forwards (i.e., $\theta$ is increasing). The second one is a flight phase, which occurs when the rigid body moves, describing a ballistic motion under the influence of gravity. The transition between both phases occurs when the spring is uncompressed (i.e., it is unloaded) until the time when the spring touch the ground again. Both phases define a hybrid system.

The Lagrangian describing the flight phase is given by the kinetic energy minus the potential energy,  which is given by the spring potential $V(\xi)$ and the gravitational potential, that is, $$L(\xi,\varphi,\theta,\dot{\xi},\dot{\varphi},\dot{\theta})=\frac{1}{2}m(\dot{\xi}^{2}+\xi^2\dot{\varphi}^{2})+\frac{1}{2}I\dot{\theta}^{2}-(mg\xi\cos\varphi+V(\xi)).$$ 

To derive the switching surface and the impact map, we note that the flight starts when the spring length reaches its non-load length (i.e., $\xi=l_0$). Therefore the switching surface is given by $\mathcal{S}=\{(\xi,\varphi,\theta,\dot{\xi},\dot{\varphi},\dot{\theta})\in TQ|\,\xi=l_0\}$.

 By assuming that at the start of the stance phase the leg is at an angle of $-\varphi_0$, by employing polar coordinates $x=\xi\sin\varphi$, $y=\xi\cos\varphi$ and nothing that $y^{+}=l_0\cos(-\varphi_0)=l_0\cos(\varphi_0)$, the impact map is given by $$\Delta(x^{-},y^{-},\theta^{-},\dot{x}^{-},\dot{y}^{-},\dot{\theta}^{-})=(-l_0\sin\varphi_0,l_0\cos\varphi_0,-\theta^{-},\dot{x}^{-},-\dot{y}^{-}, -\dot{\theta}^{-}).$$

Note that the Lagrangian and the constraint which define the switching surface, i.e., $h(q)=\xi-l_0$, are both cyclic in $\theta$. Therefore by denoting $\mu$ the conserved quantity, the Routhian $R_{c}^{\mu}:T(\mathbb{S}^{1}\times\mathbb{R})\to\mathbb{R}$ is given by $$R_{c}^{\mu}(\xi,\varphi,\dot{\xi},\dot{\varphi})=\frac{m}{2}(\dot{\xi}^{2}+\xi^{2}\dot{\varphi}^{2})-\frac{\mu^2}{2I}-mg\xi\cos\varphi-V(\xi).$$  

 Note that $R_c^{\mu}$ describes the motion for the spring-loaded inverted pendulum (SLIP) which has been used as a model which reasonably provides a template for sagittal plane motions of the center of mass (COM) of diverse legged systems as it was reviewed in  \cite{HoFuKoGu} and further studied in \cite{galiaza} and \cite{De}. 
 
\begin{figure}[h!]\label{ffig4}\centering
 \includegraphics[width=6.5cm]{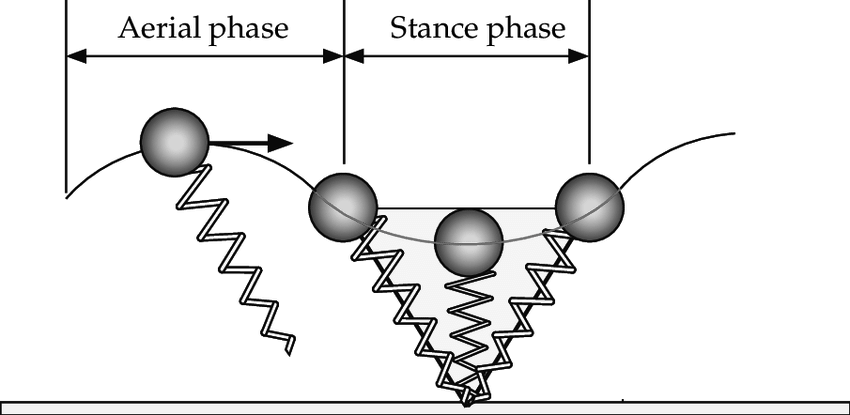}
     \caption{2D spring loaded inverted pendulum}
\end{figure}

The aerial phase consists of a projectile (or ballistic) motion  for the COM (where the only external force is
gravity) at the end of which, when $\xi = l_0$, the next stance phase starts as is shown in Figure \ref{ffig4}.

Routh equations for $R_c^{\mu}$ are given by $$
\ddot{\xi}=\xi\dot{\varphi}^2-g\cos\varphi-\frac{1}{m}\frac{\partial V}{\partial\xi},\quad
\ddot{\varphi}=\frac{g}{\xi}\sin\varphi-\frac{2\dot{\varphi}\dot{\xi}}{\xi},$$which defines the Routh vector field $$X_{R_c^{\mu}}(s)=\left(s,\dot{\xi},\dot{\varphi},\xi\dot{\varphi}^2-g\cos\varphi-\frac{1}{m}\frac{\partial V}{\partial\xi}, \frac{g}{\xi}\sin\varphi-\frac{2\dot{\varphi}\dot{\xi}}{\xi}\right)^{T},$$where $s(t)=(\xi(t),\dot{\xi}(t),\varphi(t),\dot{\varphi}(t))\in T(\mathbb{R}\times\mathbb{S}^{1})$.
Therefore the simple hybrid Routhian system $\Sigma_{\textbf{H}^{R_c^{\mu}}}$  is given by 
 \begin{equation}\label{RexHDS}\Sigma_{\mathbf{H}^{R^{\mu}_c}}:\begin{cases} \dot{s}(t)=X_{R_c^{\mu}}(s(t)), \hbox{ if } s^{-}(t)\notin\mathcal{S}^{\mu},\\ s^{+}(t)=\Delta^{\mu}(s^{-}(t)),\hbox{ if } s^-(t)\in\mathcal{S}^{\mu}, \end{cases}\end{equation}where $$\mathcal{S}^{\mu}=\{(\xi,\dot{\xi},\varphi,\dot{\varphi})\in T(\mathbb{R}\times\mathbb{S}^{1})|\xi=l_0\}$$ and $$\Delta^{\mu}(x^{-},y^{-},\dot{x}^{-},\dot{y}^{-})=(-l_0\sin\varphi_0,l_0\cos\varphi_0,\dot{x}^{-},-\dot{y}^{-}).$$\end{example}

\section{Time reversal symmetries and periodic solutions for Simple Hybrid Routhian systems}\label{sec4}
In this section we study how to impose symmetries on the Routhian vector field $X_{R_c^{\mu}}$ describing the continuous-time dynamics for a simple hybrid Routhian system, and onto the impact map, to achieve periodic motions in these classes of hybrid systems. 

\subsection{Time-reversal symmetries for simple hybrid Routhian systems}
As it was reviewed in \cite{Lamb} (see also \cite{Al}), the notion of time reversal symmetries plays a fundamental role in mechanical systems which are invariant under the transformation $(q,\dot{q},t)\mapsto (q,-\dot{q},-t)$. This symmetry implies that for a trajectory in phase space $\gamma(t)=(q(t),\dot{q}(t))$ with initial condition $\gamma_0=(q_0,\dot{q}_0)$, then $\beta(t)=(q(-t),-\dot{q}(-t))$ is also a solution for the system with initial condition $\beta_0=(q_0,-\dot{q}_0)$. In particular, if we have the trajectory $q(t)$, we also have the trajectory $q(-t)$.

\begin{definition}{\cite{Lamb}}
A diffeomorphism $\Phi:TP\to TP$ is called \textit{time-reversal symmetry} for the Routh vector field $X_{R_c^{\mu}}$ with Routhian $R_c^{\mu}:TP\to\mathbb{R}$ if $\Phi$ is an involution, that is, $\Phi\circ \Phi=Id$, and it satisfies \begin{equation}\label{reversing}\frac{d\Phi}{dt}(\gamma(t))=-X_{R_c^{\mu}}(\Phi(\gamma(t))).\end{equation} That is, the Routh vector field satisfies $X_{R_c^{\mu}}(\Phi(q,\dot{q}))=-d\Phi(q,\dot{q})\cdot X_{R_c^{\mu}}(q,\dot{q})$.

We call a Routh vector field satisfying condition \eqref{reversing} a \textit{reversible Routh vector field} under the time-reversing symmetry $\Phi$. \end{definition}

\begin{remark}
Note that the name ``time-reversal'' is given by the fact that equation \eqref{reversing} can be read as $$\Phi\circ X_{R_c^{\mu}}^{t}=X_{R_c^{\mu}}^{-t}\circ \Phi,$$ where $-t$ means the time-reversibility of the vector field $X_{R_c^{\mu}}^{t}$.
\end{remark}

\begin{proposition}
Consider a time-reversal symmetry $\Phi$ for $X_{R_{c}^{\mu}}$. If $\gamma^{*}$ is a fixed point of $\Phi$ such that $\gamma(0)=\gamma^{*}$ for $\gamma$ an integral curve of $X_{R_c^{\mu}}$ passing through $\gamma^{*}$, then $\Phi(\gamma(t))=\gamma(-t)$.
\end{proposition}

\textbf{Proof:} Consider $\tilde{\gamma}(t)=\Phi(\gamma(-t))$, then $\tilde{\gamma}(0)=\Phi(\gamma(0))=\Phi(\gamma^{*})=\gamma^{*}.$ That is, $\gamma$ and $\tilde{\gamma}$ satisfy the same initial value. 

Note that $\dot{\tilde{\gamma}}(t)=-d\Phi\cdot\dot{\gamma}(-t)$, but $\gamma(t)$ is a solution of $X_{R_c^{\mu}}$, then $\dot{\tilde{\gamma}}(t)=-d\Phi\cdot X_{R_c^{\mu}}(\gamma(-t))$.

Since $\Phi$ is a time-reversal symmetry for $X_{R_c^{\mu}}$, then $\dot{\tilde{\gamma}}(t)=X_{R_c^{\mu}}(\Phi(\gamma(-t)))$, and by definition of $\tilde{\gamma}(t)$, it follows that $\dot{\tilde{\gamma}}(t)=X_{R_c^{\mu}}(\tilde{\gamma}(t))$. Therefore, $\tilde{\gamma}(t)$ is a solution for $X_{R_c^{\mu}}$ with initial value $\gamma^{*}$. By uniqueness of solutions for an initial value problem, $\tilde{\gamma}(t)=\gamma(t)$, that is $\Phi(\gamma(-t))=\gamma(t)$. Since $\Phi$ is an involution, $\Phi(\gamma(t))=\gamma(-t)$.\hfill$\square$

\begin{theorem}\label{prop1}
Let $R_{c}^{\mu}:TP\to\mathbb{R}$ be the Routhian function invariant under the map $\Phi:TP\to TP$, \begin{equation}\label{Gsym}\Phi(q(t),\dot{q}(t))=(F(q(t)),-dF(q)\cdot \dot{q}(t))\end{equation} with $F:Q\to Q$ a smooth involution. If $\gamma^{*}=(q^{*},\dot{q}^{*})$ is a fixed point of $\Phi$, then $\Phi(\gamma(t))=\gamma(-t)$. In particular $F(q(t))=q(-t)$. 
\end{theorem}
\textbf{Proof:} Let $\gamma(t)$ be a solution of Routh equations for $R_{c}^{\mu}$ with initial value $\gamma(0)=\gamma^{*}$ and consider the map $\tilde{\gamma}(t)=\Phi(\gamma(-t))$. Given that $\gamma^{*}$ is a fixed point of $\Phi$ both curves in $TP$, $\gamma(t)$ and $\tilde{\gamma}(t)$, satisfy the same initial values. 
Since $R_{c}^{\mu}$ is invariant under $\Phi$, $R_c^{\mu}(\tilde{\gamma}(t))=R_c^{\mu}(\Phi(\gamma(-t)))=R_c^{\mu}(\gamma(-t))$ and given the Routh's equations are invariant under $(q,\dot{q},t)\mapsto(q,-\dot{q},-t)$, $R_c^{\mu}(\tilde{\gamma}(t))=R_{c}^{\mu}(\gamma(t))$. Therefore, $\tilde{\gamma}(t)$ and $\gamma(t)$ satisfy the same Routh equations. Next, by uniqueness of solutions for the initial value problem, it follows that $\Phi(\gamma(t))=\gamma(-t)$. Therefore, $F(q(t))=q(-t)$.\hfill$\square$
\subsection{Existence of periodic orbits}
In this section, based on the properties of the Routhian, we find sufficient conditions for the existence of periodic solutions in simple hybrid Routhian systems in analogy with the results for invariant Hamiltonian systems  studied in \cite{CoClBl}.
\begin{theorem}\label{prop2} Let $\Sigma_{\mathbf{H}^{R^{\mu}_c}}$ be a simple hybrid Routhian dynamical system with Routhian $R_c^{\mu}:TP\to\mathbb{R}$ invariant under $\Phi:TP\to TP$ defined as in \eqref{Gsym}. If $\gamma^{*}$ is a fixed point of $\Phi$, $\gamma$ crosses the switching surface $\mathcal{S}$ at $t_{i}^{-}=\hbox{ inf }\{t>0|\gamma(t)\in \mathcal{S}\}$ and the impact map is defined as $\Delta(\gamma^{-}(t_i))=\Phi(\gamma^{-}(t_i))$ then $\gamma(t)$ is a periodic solution for  $\Sigma_{\mathbf{H}^{R^{\mu}_c}}$ with period $2t_i^{-}$.

\end{theorem}

\textbf{Proof:} Since $R_{c}^{\mu}$ is invariant under $\Phi$, by Theorem \ref{prop1} $\Phi$ must satisfy $\Phi(\gamma(t))=\gamma(-t)$. In particular, for $t=t_i^{-}$, using the notation $\gamma(t_i^{-})=\gamma^{-}(t_i)$, we have $\Phi(\gamma^{-})=\gamma(-t_i^{-})=\gamma^{-}(-t_i)$. Given that by definition $\Delta(\gamma^{-})=\gamma^{+}(t)$ we have that $\gamma(t_i^{+})=\gamma(-t_i)$ and therefore, right after the impact, $\gamma(t)$ is re-initialized back to $\gamma(-t_i)$ so, it is periodic with period $2t_i^{-}$. 
  \hfill$\square$
  
The advantage of this result is that we can search for periodic orbits just looking at the Routhian function, instead of, for instance, using the Poincar\'e return map. Also, depending on the quantity of fixed points, Theorem \ref{prop2} provides a family of periodic solutions instead of a single periodic orbit.
\begin{example}[The 2D robotic hopper - continuation]\label{slip2} Consider the $2D$ hybrid system given by the planar robotic hopper introduced in Example \ref{slip}. We have seen that after employing Routh reduction this system becomes in the 2D SLIP. 

Consider the function $F:\mathbb{R}\times\mathbb{S}^{1}\to\mathbb{R}\times\mathbb{S}^{1}$ as $F(\xi,\varphi)=(\xi,-\varphi)$. $F$ is a smooth involution. Using $F$ we can construct the symmetry map $\Phi:T(\mathbb{R}\times\mathbb{S}^{1})\to T(\mathbb{R}\times\mathbb{S}^{1})$ using \eqref{Gsym} as $\Phi(\xi,\varphi,\dot{\xi},\dot{\varphi})=(\xi,-\varphi,-\dot{\xi},\dot{\varphi})$. It is easy to check that $R_{c}^{\mu}\circ \Phi=R_c^{\mu}$. 

Fixed points of $\Phi$ are given by $\gamma^{*}=(\xi^{*},0,0,\dot{\varphi}^{*})$, for any $\xi^{*}$ and $\dot{\varphi}^{*}$. Let $t_i^{-}$ the point where $\gamma$ crosses the switching surface $\mathcal{S}^{\mu}$ and define $\Delta^{\mu}(t_i^{-})=\Delta^{\mu}(\xi^{-},\varphi^{-},\dot{\xi}^{-},\dot{\varphi}^{-})=\Phi(\xi^{-},\varphi^{-},\dot{\xi}^{-},\dot{\varphi}^{-})=(l_0,-\varphi_0,-\dot{\xi}^{-},\dot{\varphi}^{-})$. Therefore, by   Theorem \ref{prop2}, there exists a periodic solution for the hybrid Routhian system determined by $R_c^{\mu}$ and $\Delta^{\mu}$ with period $2t_i^{-}$.
\end{example}

\subsection{Stability analysis of periodic orbits}
Let $\gamma(t)$ be a periodic solution for $\Sigma_{\mathbf{H}^{R^{\mu}_c}}$ (with period $2t_i^{-}$) associated with the time-reversal symmetry $\Phi:TP\to TP$ and $\gamma^{*}=\gamma(0)$ be a fixed point of $\Phi$. For the stability analysis of this orbit, we use the method of Poincar\'e maps \cite{p1}, \cite{p3}.

Let $\mathcal{P}$ be the Poincar\'e map corresponding to the periodic orbit $\gamma(t)$, that is, $\mathcal{P}:\mathcal{S}^{\mu}_{\gamma^{*}}\to\mathcal{S}^{\mu}_{\gamma^{*}}$, being $\mathcal{S}^{\mu}_{\gamma^{*}}$ (i.e., the reduced switching surface at the fixed point $\gamma^{*}$) the \textit{Poincar\'e section}, a hypersurface at $\gamma^{*}$, with co-dimension one of the reduced configuration space, where we are assuming that $\gamma^{*}\in\mathcal{S}_{\gamma^{*}}^{\mu}$. 

Given that $\gamma$ is a periodic orbit and $\gamma(0)=\gamma^{*}$, by definition of Poincar\'e map, $\mathcal{P}(\gamma^{*})=\gamma^{*}$. Stability analysis employing the method of Poincar\'e maps tell us that $\gamma(t)$ is asymptotically stable at $\gamma^{*}$ if the eigenvalues for the Jacobian of $\mathcal{P}$ (i.e., its tangent map) at $\gamma^{*}$, denoted by $\mathcal{T}\mathcal{P}:T_{\gamma^{*}}\mathcal{S}^{\mu}_{\gamma^{*}}\to T_{\gamma^{*}}\mathcal{S}^{\mu}_{\gamma^{*}}$, are within the unit circle (that is, if the discrete system $\gamma_{n+1}=\mathcal{P}(\gamma_n)$ is asymptotically stable at $\gamma^{*}$). 

Denote by $\hbox{Fix}(f)$ the set of fixed points associated with a function $f:TP\to TP$, that is, $$\hbox{Fix}(f)=\{x\in TP|\,f(x)=x\}.$$

Given that $\Phi$ is a diffeomorphism, $\hbox{Fix}(\Phi)$ is an embedded submanifold of $TP$ and we assume it has constant dimension, $r=\dim(\hbox{Fix}(\Phi))<\dim(TP)$.  Therefore, there exists a hypersurface $\mathcal{S}^{\mu}_{\gamma^{*}}$ at $\gamma^{*}$ such that $T_{\gamma}\hbox{Fix}(\Phi)\subset T_{\gamma}\mathcal{S}_{\gamma^{*}}^{\mu}$. We denote by $\{q^{\alpha}\}=(q^1,\ldots,q^r)$ with $1\leq\alpha\leq r$ local coordinates on the submanifold $\hbox{Fix}(\Phi)$.  Therefore, in this set of local coordinates, $\gamma^{*}\in TP$ has the expression, $\gamma^{*}=(\gamma^{*}_1,\ldots,\gamma^{*}_r,0,\ldots,0)\in\hbox{Fix}(\Phi)$, then $\mathcal{P}(\gamma^{*},0)=\gamma^{*}$.

\begin{definition}[\cite{Biped-book}]
Let $\phi(t,\alpha_0)$ be a solution for $\Sigma_{\mathbf{H}^{R^{\mu}_c}}$. The map $T_{\Delta^{\mu}}:TP\to\mathbb{R}\cup\{\infty\}$ given by \begin{equation}\label{tti}T_{\Delta^{\mu}}(\alpha_0)=\begin{cases} \hbox{inf }\{t\leq 0|\phi(t,\alpha_0)\in\mathcal{S}^{\mu}\} \hbox{ if there exists }t: \phi(t,\alpha_0)\in\mathcal{S}^{\mu}, \\ \infty \hbox{ otherwise } \end{cases}\end{equation} is called time-to-impact map.
\end{definition}

Next, we denote by $\lambda_i$ the eigenvalues of the Jacobian $\mathcal{T}\mathcal{P}$, and 
$$\Lambda_0=\{\# \lambda_i \hbox{ such that }|\lambda_i|=0\},\quad\Lambda_1=\{\# \lambda_i \hbox{ such that }|\lambda_i|=1\}$$where $\#$ means``quantity''.

\begin{lemma}
Let $\gamma(t)$ be a periodic solution for  a simple hybrid system $\Sigma_{\textbf{H}}$ as in equation \eqref{LHS}, and $\mathcal{P}:\mathcal{S}\to\mathcal{S}$ the corresponding Poincar\'e map to $\gamma(t)$. If $\hbox{rank}(\Delta)=\beta$ is constant, then $\Lambda_0\geq n-1-\beta$.
\end{lemma}

\textbf{Proof:} Consider the function $N:\Delta(\mathcal{S})\to\mathcal{S}$ given by $N(x)=\phi(T_{\Delta}(x),x)$. Therefore, $\mathcal{P}(x)=N(\Delta(x))$ and it follows that $$\mathcal{T}\mathcal{P}(\gamma^{*})=dN(\Delta(\gamma^{*}))\cdot d\Delta(\gamma^{*}).$$

Given that $\hbox{rank } dN(\Delta(\gamma^{*}))\cdot d\Delta(\gamma^{*})\leq\hbox{ rank }(d\Delta(\gamma^{*}))=\beta$ and $\dim(\mathcal{S})\leq n-1$, by rank-nullity Theorem and  the fact that kernel of $\mathcal{T}\mathcal{P}$ is precisely the eigenspace corresponding to the eigenvalue $0$, if follows that $\Lambda_0\geq n-1-\beta$.\hfill$\square$. 

\begin{theorem}\label{thpm}
Let $\Sigma_{\mathbf{H}^{R^{\mu}_c}}$ be a simple hybrid Routhian system satisfying $\Phi(\gamma(t))=\gamma(-t)$ with $\gamma(0)=\gamma^{*}$ a fixed point of $\Phi$. If $\gamma(t)$ is a periodic solution transversal to $\mathcal{S}_{\gamma^{*}}^{\mu}$ at $\gamma^{*}$,  then $\Lambda_1\geq r$.
\end{theorem}
\textbf{Proof:} By the transversality assumption, we can employ [Theorem $3.3$ in \cite{GoCo}] and so there is an open subset $\mathcal{O}\subset TP$ of $\gamma^{*}$ with $\mathcal{S}_{\gamma^{*}}^{\mu}\subset \mathcal{O}$, where every trajectory starting form $\mathcal{O}$ crosses $\mathcal{S}^{\mu}$ and where there exists a Poincar\'e map $\mathcal{P}:S_{\gamma^{*}}^{\mu}\to S_{\gamma^{*}}^{\mu}$. Denote by 
$\mathcal{P}=[\mathcal{P}_1,\mathcal{P}_2,\ldots,\mathcal{P}_{n-1}]$ the Poincar\'e map in local coordinates $(x^a,\dot{x}^{a})\in TP$, satisfying $\mathcal{P}(\gamma^{*},0)=\gamma^{*}$ where $\gamma^{*}\in\hbox{Fix}(\Phi)$. Since $\Phi$ is a time-reversal symmetry, every solution starting at $\gamma^{*}$ is a periodic orbit.

It is easy to check that $\mathcal{T}\mathcal{P}_{ij}(\gamma^{*})=\delta_{ij}$ with $\delta_{ij}=1$ if $i=j$ and $\delta_{ij}=0$ if $i\neq j$, for $i=1,\ldots,n-1$ and $j=1,\ldots,r$, where $\mathcal{T}\mathcal{P}_{ij}$ denotes the $(i,j)$-entries for the Jacobian matrix of $\mathcal{P}$. Indeed, it follows from 

\begin{align*}\mathcal{T}\mathcal{P}_{ii}(\gamma^{*})=&\displaystyle{\lim_{h\to0}}\frac{\mathcal{P}_{i}(\gamma_1^{*},\ldots,\gamma_i^{*}+h,\ldots,\gamma_k^{*},0,\ldots,0)-\mathcal{P}_{i}(\gamma_1^{*},\ldots,\gamma_i^{*},\ldots,\gamma_k^{*},0,\ldots,0)}{h}\\=&\displaystyle{\lim_{h\to 0}}\frac{\gamma_i^{*}+h-\gamma_i^{*}}{h}=1,\end{align*}and the fact that if $i\neq j$, 

\begin{align*}\mathcal{T}\mathcal{P}_{ij}(\gamma^{*})=&\displaystyle{\lim_{h\to0}}\frac{\mathcal{P}_{i}(\gamma_1^{*},\ldots,\gamma_j^{*}+h,\ldots,\gamma_k^{*},0,\ldots,0)-\mathcal{P}_{i}(\gamma_1^{*},\ldots,\gamma_i^{*},\ldots,\gamma_k^{*},0,\ldots,0)}{h}\\
=&\displaystyle{\lim_{h\to 0}}\frac{\gamma_i^{*}-\gamma_i^{*}}{h}=0.\end{align*}Therefore $\mathcal{T}\mathcal{P}$ has at least $r$ eigenvalues $\lambda_i$ with $|\lambda_i|=1$.\hfill$\square$

\begin{example}[The 2D robotic hopper - continuation]\label{ex4}
Consider the situation of Example \ref{slip2}, which is a $4$-dimensional system on $T(\mathbb{R}\times\mathbb{S}^{1})$. The fixed points of $\Phi$ are $\gamma^{*}=(\xi^{*},0,0,\dot{\varphi}^{*})$. Then $\dim(\hbox{Fix}(\Phi))=2$.

By Theorem \ref{thpm}, $\Lambda_1\geq2$. Then $\mathcal{T}\mathcal{P}(\gamma^{*})$ is of the form 
\[
   \mathcal{T}\mathcal{P}(\gamma^{*})=
  \left[ {\begin{array}{cccc}
   1 & 0 & \ast & \ast \\
   0 & 1 & \ast & \ast\\
   0 & 0 & \ast & \ast\\
   0 & 0 & \ast & \ast\\
  \end{array} }\right].
\] Given that $\hbox{rank}({\Delta^{\mu}})=2$, $\Lambda_0\geq 1$. Therefore $\mathcal{T}\mathcal{P}(\gamma^{*})$, in an appropriate choice of coordinates, takes the form \[
   \mathcal{T}\mathcal{P}(\gamma^{*})=
  \left[ {\begin{array}{cccc}
   1 & 0 & \ast & 0 \\
   0 & 1 & \ast & 0\\
   0 & 0 & \ast & 0\\
   0 & 0 & \ast & 0\\
  \end{array} }\right].
\]  

It follows that the set of eigenvalues for $\mathcal{T}\mathcal{P}$ are $\{1,1,\lambda,0\}$. Hence, if $|\lambda|<1$, the periodic orbit is marginally stable. 
\end{example}

\begin{remark}
Note that in Example \ref{ex4} we can get, at most a characterization of neutral stability for the periodic solution $\gamma$. It would be interesting to consider perturbed simple hybrid Routhian systems, similarly to the framework given in \cite{hamed}, while the perturbation preserves the symmetry, in order that we can turn the neutrally stable periodic orbit into a stable limit cycle. This will be explored in a future work by considering an adaptation of the averaging method and approximate for hybrid systems given in \cite{De} and \cite{tabuada1}, respectively.
\end{remark}

\section{Application to existence of periodic solutions for hybrid Routhian control systems}

In this section we apply the results given in Section \ref{sec4} to underactuated control systems. We study how by considering the notion of hybrid zero dynamics given \cite{Biped-book} together with a time reversible symmetry we can obtain a characterization which facilitates the searching of periodic solutions in simple hybrid Routhian control system.

\subsection{Underactuated mechanical control system}

An underactuated control system is a control system where the quantity of actuators is fewer than the dimension of the configuration space. 

Consider a Lagrangian function $L:TQ  \rightarrow
\mathbb{R}$, with $\dim(Q)=n$ which is cyclic with respect to one of the generalized coordinates. Without loss of generality, we assume that the cyclic variable is the last one, that is, $q^{i}=(x^{a},\theta)$, $a=1,\ldots,n-1$, with $\theta$ cyclic. In order to design control laws for controlled simple hybrid Routhian systems we consider the \textit{underactuated controlled Euler-Lagrange equations} \cite{Bl}
\begin{align*}\frac{d}{dt}\left(\frac{\partial L}{\partial\dot{x}^{\alpha}}\right) -
\frac{\partial L}{\partial x^{\alpha}}&=u_{\alpha},\quad \frac{d}{dt}\left(\frac{\partial L}{\partial\dot{\theta}}\right) -
\frac{\partial L}{\partial \theta}=0,\\
\frac{d}{dt}\left(\frac{\partial L}{\partial\dot{x}^{\beta}}\right) -
\frac{\partial L}{\partial x^{\beta}}&=0,
\end{align*}
where $\alpha=1,\ldots,k$; $\beta=k+1,\ldots, n-1$, with $u(t)=(u_1(t),...,u_{k}(t))\in U$ control inputs and 
where  $U$ is an open subset of $\mathbb{R}^{k}$, the set of admissible controls. 

We assume at least one degree of underactuation, where the underactuated configurations include the cyclic variable $\theta$.

Since the cyclic variable is uncontrolled, by employing Routh reduction, the reduced equations are given by the controlled Routh equations for the Routhian $R_c^{\mu}:TP\to \mathbb{R}$, that is, \begin{equation}\label{controlR}
\frac{d}{dt}\left(\frac{\partial R_c^{\mu}}{\partial\dot{x}^{a}}\right)-\frac{\partial R_c^{\mu}}{\partial x^{a}}=u_{\alpha},\quad
\frac{d}{dt}\left(\frac{\partial R_c^{\mu}}{\partial\dot{x}^{\beta}}\right) -
\frac{\partial R_c^{\mu}}{\partial x^{\beta}}=0.\end{equation}
Equations \eqref{controlR} give rise to a model of an affine control system of the form 
\begin{equation}\label{vfHC}
\dot{\gamma}=X_{R_c^{\mu}}(\gamma)+C(\gamma)u:=X(\gamma,u)\end{equation}
where  $C$ is a constant matrix, $X_{R_c^{\mu}}$ is the Routh vector field and $X:TP\times U\to T(TP)$ is a vector field, called control vector field.

The tuple $\mathbf{H}_{\mathbf{c}}^{R_c^{\mu}}=(TP, U, \mathcal{S}^{\mu}, \Delta^{\mu}, X)$ with $\mathcal{S}^{\mu}$ and $\Delta^{\mu}$ as in Definition \ref{def1}, and $X:TP\times U\to T(TP)$ defined in \eqref{vfHC}, is called \textit{simple hybrid Routhian control system}.
A simple hybrid Routhian system is a simple hybrid Routhian control system with $U=\{0\}$.

\subsection{Hybrid zero dynamics and periodic solutions for simple hybrid\\ Routhian control systems}

\begin{definition}[\cite{Isidori}]\label{def4.2}
Consider the control vector field $X(\gamma,w)$ given in \eqref{vfHC}. The embedded submanifold $\mathcal{Z}$ of $TP$ given by $$\mathcal{Z}=\{\tilde{\gamma}\in TP|\, \exists! u^{\star}(\tilde{\gamma})\hbox{ s.t. } X(\tilde{\gamma},u^{\star}(\gamma))\in T_{\tilde{\gamma}}\mathcal{Z}\}$$ is called the \textit{zero dynamics submanifold} of $TP$, and $\dot{\tilde{\gamma}}=X(\tilde{\gamma},u^{\star}(\tilde{\gamma}))$ is the associated \textit{zero dynamics} on $\mathcal{Z}$.
\end{definition}

\begin{theorem}\label{theorem4.2}
Consider a control system $\dot{\gamma}=X(\gamma,u)$ as in \eqref{vfHC} with associated Routhian vector field $X_{R_c^{\mu}}$ and $\mathcal{Z}$ the associated zero dynamics submanifold. Assume that $\mathcal{Z}$ is invariant under the time reversal symmetry $\Phi$ for $X_{R_c^{\mu}}$, \begin{equation}\label{gammaeq}C(\Phi(\gamma))\Gamma(u)=-(d\Phi(\gamma)C(\gamma))u\end{equation} for all $\gamma\in TP$ and with $\Gamma:U\to U$ a one to one invertible map. If $\gamma^{*}$ is a fixed point of $\Phi$ in $\mathcal{Z}$, the solution $\gamma(t):I\to TP$ with initial condition $\gamma(0)=\gamma^{*}$ belongs to $\mathcal{Z}$ and $\Phi(\gamma(t))=\gamma(-t)$ $\forall t\in I$. Moreover, for all $\gamma\in\mathcal{Z}$ it follows that $u^{\star}(\Phi(\gamma))=\Gamma(u^{\star}(\gamma))$ and $$\mathcal{X}(\Psi(\gamma))=-d\Psi(\gamma)\cdot \mathcal{X},$$ where $\mathcal{X}$ and  $\Psi$ denotes the restrictions to $\mathcal{Z}$ of $X$ and $\Phi$ respectively. \end{theorem}

\textbf{Proof:}  If $\gamma^{*}\in\mathcal{Z}$ is a fixed point of $\Phi$, by definition of the zero dynamics $X(\gamma,u^{\star}(\gamma))\in T_{\gamma}\mathcal{Z}$ $\forall\gamma\in\mathcal{Z}$, hence $\gamma(t)\in\mathcal{Z}$ $\forall t\in I$. Moreover, by Theorem \ref{prop1} $\Phi(\gamma(t))=\gamma(-t)$. 

Now, given that $\mathcal{Z}$ is invariant under $\Phi$, that is $\Phi(\gamma)\in\mathcal{Z}$ $\forall\gamma\in\mathcal{Z}$, then \begin{equation}\label{eqq1}-d\Phi(\gamma)\cdot X(\gamma,u^{\star}(\gamma))\in T_{\Phi(\gamma)}\mathcal{Z}.\end{equation}

 By the definition of the zero dynamics $X(\Phi(\gamma),u^{\star}(\Phi(\gamma))) \in T_{\Phi(\gamma)}\mathcal{Z}.$ Using that $C(\Phi(\gamma)))\Gamma(u(\gamma))=-(d\Phi(\gamma)C(\gamma))u(\gamma)$ and the fact that $\Phi$ is a time reversible symmetry for $X_{R_c^{\mu}}$, $-d\Phi(\gamma)\cdot X_{R_c^{\mu}}(\gamma)=X_{R_c^{\mu}}(\Phi(\gamma))$. Therefore, $$X_{R_c^{\mu}}(\Phi(\gamma))+C(\Phi(\gamma))\Gamma(u^{\star}(\gamma))\in T_{\Phi(\gamma)}\mathcal{Z}.$$ 
 
 Finally, by Definition \ref{def4.2} the feedback control $u^{\star}(\gamma)$ is unique, then \eqref{eqq1} is equivalent to $X_{R_c^{\mu}}(\Phi(\gamma))+C(\Phi(\gamma))\Gamma(u^{\star}(\gamma))$. Therefore $\forall\gamma\in\mathcal{Z}$, $\Gamma(u^{\star}(\gamma))=u^{\star}(\Phi(\gamma))$ and given that $\Phi$ is a time reversible symmetry for $X_{R_c^{\mu}}$, it follows that $-d\Phi(\gamma)\cdot X(\gamma,u^{\star}(\gamma))=X(\Phi(\gamma),u^{\star}(\Phi(\gamma)))\in T_{\Phi(\gamma)}\mathcal{Z}$ .\hfill$\square$

The following definition are given in analogy with \cite{Biped-book} for the class of simple hybrid Routhian control systems:

\begin{definition}
Let $\mathbf{H}_{\mathbf{c}}^{R_c^{\mu}}$ be a simple hybrid Routhian control system and let $\mathcal{Z}$ be the zero dynamics submanifold for $X_{R_c^{\mu}}$ imposed by $u^{\star}(\gamma)$. Denoting by $\mathcal{W}=\mathcal{Z}\cap\mathcal{S}^{\mu}$, the submanifold $\mathcal{Z}$ is called \textit{hybrid invariant} if $\Delta^{\mu}(\mathcal{W})\subset\mathcal{Z}$. 
\end{definition}
\begin{definition}
Consider the simple hybrid Routhian control system $\widetilde{\mathbf{H}}_{\mathbf{c}}^{R_c^{\mu}}=(\mathcal{Z},U,\mathcal{W},\Delta^{\mu}|_{\mathcal{Z}},\mathcal{X})$ generating the hybrid dynamical control system \begin{equation}\label{HDCS}\Sigma_{\widetilde{\textbf{H}}_{\mathbf{c}}^{R_c^{\mu}}}:\begin{cases} \dot{\gamma}(t)=\mathcal{X}(\gamma(t),u^{\star}(\gamma(t))),\quad\gamma^{-}(t)\notin\mathcal{X} \\ \gamma^{+}(t)=\Delta^{\mu}|_{\mathcal{Z}}(\gamma^{-}(t)),\quad\quad \gamma^-(t)\in\mathcal{W}. \end{cases}\end{equation} The hybrid dynamical control system $\Sigma_{\widetilde{\textbf{H}}_{\mathbf{c}}^{R_c^{\mu}}}$ is called \textit{hybrid zero dynamics} associated with the simple hybrid Routhian control system $\textbf{H}_{\mathbf{c}}^{R_c^{\mu}}$.
\end{definition}

\begin{theorem}
Consider the situation and hypothesis of Theorem \ref{theorem4.2}, but where $\gamma$ is any solution of $\Sigma_{\widetilde{\textbf{H}}_{\mathbf{c}}^{R_c^{\mu}}}$ satisfying $\gamma(0)=\gamma^{*}$ with $\gamma^{*}\in\mathcal{Z}$ a fixed point of the time reversal symmetry $\Phi$ for $X_{R_c^{\mu}}$ with $\mathcal{Z}$ hybrid invariant. If in addition $\gamma$ crosses the switching surface $\mathcal{W}$ at $t_{i}^{-}=\hbox{inf }\{t>t_{i-1}|\gamma(t)\in \mathcal{W}\}$ and the impact map is defined as $\Delta^{\mu}(\gamma^{-}(t_i))=\Phi(\gamma^{-}(t_i))$, then $\gamma(t)$ is a periodic solution on $\mathcal{Z}$ for the simple hybrid Routhian control system $\Sigma_{\widetilde{\textbf{H}}_{\mathbf{c}}^{R_c^{\mu}}}$ with period $2t_i^{-}$.
\end{theorem}
\textbf{Proof:} By Theorem \ref{theorem4.2} $\gamma(t)\in\mathcal{Z}$ and $\Phi(\gamma(t))=\gamma(-t)$ $\forall t\in I$. In particular, for $t_i=t_i^{-}$, denoting $\gamma(t_i^{-})=\gamma^{-}(t_i)\in\mathcal{W}$, we have $\Phi(\gamma^{-})=\gamma^{-}(-t_i)$. Since  $\Delta^{\mu}(\gamma^{-}(t_i))=\Phi(\gamma^{-}(t_i))$, $\Delta^{\mu}(\gamma^{-})=\gamma^{+}(t)$ and then $\gamma(t_i^{+})=\gamma(-t_i)$. Therefore, right after the impact, the solution $\gamma(t)$ is re-initialized back to $\gamma(-t_i)$ so, it is periodic with period $2t_{i}^{-}.$ \hfill$\square$

Note that from the previous result, depending on the quantity of fixed points, we are able to find a family of periodic solutions, instead of a single periodic orbit.

\begin{example}[2D controlled spring loaded inverted pendulum]

Consider the 2D robotic hopper with two degrees of underactuation, one in the pitch angle and the other in the rigid body attitude (i.e., we only control the length of the spring). By employing the results given in Example \ref{slip} the controlled Routh equations are given by  $$
\ddot{\xi}=\xi\dot{\varphi}^2-g\cos\varphi-\frac{1}{m}\frac{\partial V}{\partial\xi}+u,\quad
\ddot{\varphi}=\frac{g}{\xi}\sin\varphi-\frac{2\dot{\varphi}\dot{\xi}}{\xi}$$where $u$ is a torque applied to control the length of the spring. We consider the elastic potential to be $V(\xi)=\frac{1}{2}\kappa(\xi-l_0)^{2}$, with $\kappa\in\mathbb{R}^{+}$ the spring constant. 

Following Example \ref{slip2}, we consider the smooth involution $F(\xi(t),\varphi(t))=(\xi(t),-\varphi(t))$ and $\Phi(\gamma(t))=(\xi(t),-\varphi(t),-\dot{\xi}(t),\dot{\varphi}(t))$. Fixed points of $\Phi$ are $\gamma^{*}=(\xi^{*},0,0,\dot{\varphi}^{*})$. The Routhian $R_{c}^{\mu}$ is invariant under $\Phi$ and $X_{R_{c}^{\mu}}$ is a time reversible Routhian vector field. Then $\Phi(\gamma(t))=(\xi(-t),-\varphi(t),-\dot{\xi}(t),\dot{\varphi}(-t))$, as a consequence of Theorem \ref{prop1}. 

Next, define the zero dynamics submanifold as $$\mathcal{Z}=\left\{(\xi,\varphi,\dot{\xi},\dot{\varphi})\in T(\mathbb{R}\times\mathbb{S}^{1})|\,\xi=h(\varphi),\,\dot{\xi}=\frac{\partial h}{\partial \varphi}\dot{\varphi}\right\}$$ where $h$ is an even function of $\varphi$. Note that this choice of $h$ makes $\mathcal{Z}$ invariant under $\Phi$ and therefore the zero dynamics reads $$\ddot{\varphi}=\frac{1}{h(\varphi)}\left(g\sin\varphi-2\dot{\varphi}^{2}\frac{\partial h}{\partial\varphi}\right).$$ Note that $C(\gamma)\circ \Phi(\gamma)=-d\Phi(\gamma)C(\gamma)$ and then $\Gamma(u)=u$.

The control input $u^{\star}(\varphi,\dot{\varphi})$ on $\mathcal{Z}$ is $$
u^{\star}(\varphi,\dot{\varphi})=\frac{\partial^2h}{\partial \varphi^2}+\frac{\partial h}{\partial\varphi}\frac{g\sin\varphi}{h(\varphi)}-2\frac{\dot{\varphi}^2}{h(\varphi)}\left(\frac{\partial h}{\partial\varphi}\right)^2-h(\varphi)\dot{\varphi}^2+g\cos\varphi+\frac{\kappa(h(\varphi)-l_0)}{m}$$ and it is easy to verify that it satisfies $u^{\star}(-\varphi,\dot{\varphi})=u^{\star}(\varphi,\dot{\varphi})$, that is, $u^{\star}(\Phi(\gamma))=\Gamma(u^{\star}(\gamma))$. 

The function $\Phi$ in $\mathcal{Z}$ is given by $$\Phi(\varphi,\dot{\varphi})=(h(\varphi),-\varphi,-\frac{\partial h}{\partial\varphi}\dot{\varphi},\dot{\varphi}).$$ Since $\widetilde{\gamma}^{*}=(h(0),0,-\frac{\partial h}{\partial \varphi}|_{\varphi=0}\cdot\dot{\varphi}^{*},\dot{\varphi}^{*})$ is a fixed point of $\Phi$ in $\mathcal{Z}$, the solution $\gamma(t)$ with $\gamma(0)=\widetilde{\gamma}^{*}$ belongs to $\mathcal{Z}$. 

Finally, if we define the switching surface $\mathcal{S}$ as in Example \ref{slip2}, and impact map as  $\Delta^{\mu}(\xi^{-},\varphi^{-},\dot{\xi}^{-},\dot{\varphi}^{-})=\Phi(\xi^{-},\varphi^{-},\dot{\xi}^{-},\dot{\varphi}^{-})=(l_0,-\varphi_0,-\dot{\xi}^{-},\dot{\varphi}^{-})$ then $\mathcal{W}=\left\{(\varphi,\dot{\varphi})\in\mathcal{Z}|\,h(\varphi)=l_0\right\}$ and $\mathcal{Z}$ is hybrid invariant because $h$ is an even function. Therefore, $\gamma(t)$ is a periodic solution for the simple hybrid Routhian control system.

\end{example}

\end{document}